\documentclass[10pt]{article}

\usepackage{times}
\usepackage{amssymb}
\usepackage{prooftree}
\usepackage{amsmath}
\usepackage{latexsym}
\usepackage{stmaryrd}
\usepackage[dvips]{graphics}
\usepackage{graphicx}

\usepackage{amsmath}
\usepackage{amssymb}
\usepackage{amscd}

\usepackage[english]{babel}

\usepackage[raiselinks=false,plainpages=false,pdfpagelabels,naturalnames=true,colorlinks=true,citecolor=blue,urlcolor=blue,linkcolor=blue,bookmarksopen=true,pagebackref,hyperindex]{hyperref}

\allowdisplaybreaks[1]

\usepackage{latexsym}
\usepackage[emu]{cmll}
\usepackage{prooftree}

\def\Section#1{\section{#1}}
\def\Subsection#1{\subsection{#1}}

\def\Hide#1{\relax}

\DeclareSymbolFont{AMSb}{U}{msb}{m}{n}
\DeclareSymbolFontAlphabet{\mathbb}{AMSb}
\DeclareSymbolFont{symbolsC}{U}{txsyc}{m}{n}
\DeclareMathSymbol{\rJoin}{\mathrel}{symbolsC}{89}

\newcommand{\CK}{\vee}
\newcommand{\oCK}{\downarrow}
\newcommand{\CM}{\Rightarrow}
\newcommand{\CC}{\wedge}

\newcommand{\CKb}[2]{#1 \CK #2}
\newcommand{\oCKb}[2]{#1\downarrow{#2}}
\newcommand{\CMb}[2]{#1\Rightarrow{#2}}
\newcommand{\CCb}[2]{#1 \CC #2}

\def\proves{\vdash}

\def\Func#1{{\sf{#1}}}

\def\iAnd{\otimes} % intuitionistic conjunction
\newtheorem{Theorem}{Theorem}
\newtheorem{Lemma}[Theorem]{Lemma}
\newtheorem{Corollary}[Theorem]{Corollary}

\newtheorem{Definition}{Definition}
\newtheorem{Remark}{Remark}
\newtheorem{Example}{Example}
\def\Proof{\par \noindent{\bf Proof: }}
\def\Done{\hfill\rule{0.5em}{0.5em}}
\def\Var{\Func{Var}}

\def\lL{{\cal L}}

\def\ASM{[\Func{ASM}]}

\def\CE{[{\iAnd}\Func{E}]}
\def\CI{[{\iAnd}\Func{I}]}

\def\CONR{[\Func{CON}_r]}

\def\CON{[\Func{CON}]}

\def\eCWC{\Func{CWC}}
\def\CWC{[\eCWC]}

\def\DNE{[\Func{DNE}]}

\def\EFQ{[\Func{EFQ}]}
\def\EFQ{[\Func{EFQ}]}

\def\eLE{{\Lolly}\Func{E}}
\def\LE{[\eLE]}
\def\LI{[{\Lolly}\Func{I}]}

\newcommand{\Lnot}[1]{\neg {#1}}
\def\Lolly{\to}

\def\WK{[\Func{WK}]}

\newcommand{\Equiv}[1]{\;\leftrightarrow_{#1}\;}

\def\F{\perp}
\def\T{\top}

\def\Logic#1#2{\mbox{{\bf #1}}_{\mbox{\bf #2}}}

\def\ALi{\Logic{AL}{i}}

\def\BL{\Logic{CL}{}}
\def\ALc{\Logic{AL}{c}}

\def\LLi{\Logic{{\L}L}{i}}

\def\LLc{\Logic{{\L}L}{c}}
\def\CL{\Logic{CL}{\relax}}
\def\IL{\Logic{IL}{}}

\newcommand{\using}[1]{\mbox{[#1]}}

\newcommand{\by}[1]{\tag*{(#1)}}

\newcommand{\kTrans}[1]{{#1}^{\sf K}}

\newcommand{\Negative}{{\cal N}}

\newcommand{\glTrans}[1]{{#1}^{\sf Gli}}
\newcommand{\godTrans}[1]{{#1}^{\mbox{{\scriptsize \sf G\"{o}}}}}
\newcommand{\godTransPre}[1]{{#1}^*}
\newcommand{\ggTrans}[1]{{#1}^{\sf Gen}}
\newcommand{\krTrans}[1]{{#1}^{\mbox{{\scriptsize \sf Kr}}}}
\newcommand{\krTransPre}[1]{{#1}^\dagger}

\begin{document}

\title{Negative Translations for Affine and {\L}ukasiewicz Logic}

\author{Rob Arthan and Paulo Oliva}

\date{(\today)}

\maketitle

\begin{abstract} We investigate four well-known negative translations of classical logic into intuitionistic logic within a substructural setting. We find that in \emph{affine logic} the translation schemes due to Kolmogorov and G\"odel both satisfy Troelstra's criteria for a negative translation. On the other hand, the schemes of Glivenko and Gentzen both fail for affine logic, but for different reasons: one can extend affine logic to make Glivenko work and Gentzen fail and {\it vice versa}. By contrast,  in the setting of \emph{{\L}ukasiewicz logic}, we can prove a general result asserting that a wide class of formula translations including those of Kolmogorov, G\"odel, Gentzen and Glivenko not only satisfy Troelstra's criteria with respect to a natural intuitionistic fragment of {\L}ukasiewicz logic but are all equivalent.  
\end{abstract}

\tableofcontents

%%%%%%%%%%%%%%%%%%%%%%%%%%%%%%%%%%%%
%%%%%%%%%%%%%%%%%%%%%%%%%%%%%%%%%%%%
\Section{Introduction}
\label{sec:introduction}
%%%%%%%%%%%%%%%%%%%%%%%%%%%%%%%%%%%%
%%%%%%%%%%%%%%%%%%%%%%%%%%%%%%%%%%%%

Negative translations (also known as double negation translations) have a long history in logic and proof theory. Kolmogorov \cite{Kolmogorov(25)} was probably the first one to observe that classical logic can be ``embedded" into its intuitionistic fragment. He defined a translation $A \mapsto \kTrans{A}$ which places double negations in front of every subformula, and showed that $A$ is provable classically if and only if $\kTrans{A}$ is provable intuitionistically. Around the same time, Glivenko \cite{Glivenko(29)}, G\"odel \cite{Goedel(33)} and Gentzen \cite{Gentzen(33)} defined more ``economic" translations that also eliminate classical principles from proofs at the cost of introducing extra negations, but not as many as Kolmogorov's. 

In the present paper we recast these negative translations in the setting of substructural logic, concentrating on logics lying between intuitionistic  affine $\ALi$ logic and classical {\L}ukasiewicz logic $\LLc$. This will shed light on the amount of contraction required in order to make the translations work.

In Section \ref{sec:logics-and-algebras}, we define a fragment of classical {\L}ukasiewicz logic $\LLc$, which we will call \emph{intuitionistic {\L}ukasiewicz logic}\footnote{Our reasons for adopting this terminology are given in Section \ref{final-section}.} $\LLi$. Just as {\L}ukasiewicz logic \cite{Hajek98} is a subsystem of classical logic $\CL$, intuitionistic {\L}ukasiewicz logic is a subsystem of the usual intuitionistic logic $\IL$ \cite{Troelstra(96)}. This paper focuses on propositional logic, leaving a similar study for predicate logic to future work.

\begin{figure}[t]
\[
\begin{CD}
\ALc  @>>> \LLc  @>>> \BL \\
@AAA       @AAA       @AAA \\
\ALi  @>>> \LLi  @>>> \IL \\
\end{CD}
\]
\caption{Relationships between the six logics}
\label{fig:logics}
\end{figure} 

$\LLi$ and $\LLc$ are defined here as extensions of the $\{\Lolly, \iAnd,\F\}$-fragment of \emph{intuitionistic affine logic} $\ALi$, i.e. intuitionistic multiplicative linear logic \cite{Abramsky(93),Benton(1993),Girard(87B)} extended by allowing weakening. A similar sequent calculus for {\L}ukasiewicz logic based on classical affine logic has been proposed in \cite{Ciabattoni:1997}. The main differences are that we work on the implication-conjunction fragment of affine logic, and take \emph{intuitionistic} affine logic as the starting point.

% As it will be clear, we also focus on the intuitionistic fragments of {\L}ukasiewicz logic, and aim to prove theorems \emph{of} these system, rather than theorems \emph{about} the system (meta-theorems).
%
Starting from $\ALi$ one obtains \emph{intuitionistic {\L}ukasiewicz logic} $\LLi$ by adjoining the axiom that we call \emph{commutativity of weak conjunction} $\CWC$
\[ A \iAnd (A \Lolly B) \vdash B \iAnd (B \Lolly A) \]
which is a simple consequence of contraction, but is strictly weaker
than it. The reason we call $A \iAnd (A \Lolly B)$ a weak form of
conjunction can be explained as follows: Note that $A \iAnd (A \Lolly
B)$ implies both $A$ and $B$, but without contraction (i.e, if $A$
cannot be used twice), we do not have in general $A \iAnd (A \Lolly B)
\vdash A \iAnd B$. On the other hand, due to the presence of weakening
in the affine systems, we always have $A \iAnd B \vdash A \iAnd (A
\Lolly B)$. Hence, $A \iAnd (A \Lolly B)$ is a form of conjunction that,
in general, is strictly weaker than the usual multiplicative conjunction $A \iAnd B$. The axiom states that this conjunction is commutative. It is also known as the \emph{axiom of divisibility} in the basic logic literature \cite{jipsen-montagna06}.  

The logic $\LLi$ has been studied before, under different names. For instance, Blok and Ferreirim \cite{blok-ferreirim00} refer to it as ${\sf S}_{\cal HO}$. $\LLi$ can also be viewed as a fragment of H\'ajek's \emph{basic logic} without the (intuitionistically unacceptable) \emph{axiom of pre-linearity} \cite{Hajek98}
\[ (A \Lolly B) \Lolly C, (B \Lolly A) \Lolly C \vdash C \]
The relationship between the various logical systems is depicted as a commutative diagram in Figure \ref{fig:logics}, where  arrows indicate inclusion.

The intuitionistic systems $\ALi$ and $\LLi$ have ``classical" counterparts ($\ALc$ and $\LLc$, respectively) obtained by adding the law of \emph{double negation elimination} $\DNE$
\[ \Lnot{\Lnot{A}} \vdash A \]
where $\Lnot{A}$ is defined as $A \Lolly~\F$. 

In order to move horizontally in the diagram of Figure \ref{fig:logics} from the left-most column (affine system) to the right-most column (intuitionistic $\IL$, and classical logic $\BL$) one adds the \emph{contraction axiom} $\CON$
\[ A \vdash A \iAnd A \]
Since, over $\ALi$, $\CON$ entails $\CWC$, the {\L}ukasiewicz  systems sit in between the affine systems, where no contraction is permitted, and the systems with full contraction. In this sense, one can think of $\CWC$ as extending $\ALi$ with a restricted form of contraction that keeps track of what is left unconsumed when one uses $A$ and $A \Lolly B$ to obtain $B$.

% We give several applications of the our results, including: (1) A simpler and more abstract version of a proof, originally found by Veroff and Spinks \cite{Veroff01}, of one of Ferreirim's theorem \cite{Ferreirim92} (cf. Section \ref{sec:f-v-s-theorem}), (2) a proof that the idempotent elements of a hoop form a subhoop (cf. Section \ref{sec:idem}), (3) homomorphism properties for the double negation operator in intuitionistic {\L}ukasiewicz logic (cf. Sections \ref{homo-dn-lolly} and \ref{homo-dn-tensor}), and (4) a collection of De Morgan properties for intuitionistic {\L}ukasiewicz logic (cf. Section \ref{sec-de-morgan}). 

The main result in this paper is that all four standard negative translations of $\CL$ into $\IL$ are also negative translations of $\LLc$ into $\LLi$ (Section \ref{sec:embedding}). Our result relies on several derivations of novel theorems of $\LLi$, in particular the result that, over $\LLi$, the double negation mapping $A \mapsto \Lnot{\Lnot{A}}$ is a homomorphism (Section~\ref{sec:lli}).

We also prove that Kolmogorov's and G\"odel's translations are even
negative translations of $\ALc$ into $\ALi$ (Section
\ref{sec:neg-trans}), and, quoting \cite{arthan-oliva-2021}, briefly describe counter-examples demonstrating that Glivenko and Gentzen are not: in fact $\ALi$ can be extended so as to make the Glivenko translation a negative translation but not the Gentzen translation or {\it vice versa}.

%\begin{Remark} Another intuitionistic fuzzy logic worth mentioning is \cite{takeuti84}, although they consider \emph{extensions} of $\IL$ rather than \emph{fragments}. We have recently come across the abstract \cite{marra2014} announcing an interpretation of $\LLc$ into $\IL$. Although the short abstract does not give too much detail into the translation, it is conceivable that their interpretation will be based on a form of negative translation such as the one presented here. As they point out, however, their approach is semantic, and their result is ``purely existential". We present here a concrete fragment of $\IL$ (namely $\LLi$) and several syntactic negative translations of $\LLc$ into $\LLi$.
%\end{Remark}

In the present paper, whenever we need to show that a formula is provable in one of our logics, we do so constructively.
In the case of $\LLi$ most non-trivial derivations involve intricate applications of $\CWC$.
We express here our gratitude to the late Bill McCune for the development of the automated theorem prover
Prover9 and the finite-model finder Mace4 \cite{prover9-mace4}, which we have used extensively to find
derivations or counter-models to our various conjectures. Most of the
$\LLi$ derivations presented here were initially found by Prover9.
Perhaps remarkably, we found it possible to organise and present
the derivations in what we believe is a systematic and human-readable
style. This involved breaking down Prover9's monolithic proofs into smaller and natural lemmas of independent interest, with readable proofs that make use of previous smaller lemmas and leave easy-steps as exercises to the reader.

%%%%%%%%%%%%%%%%%%%%%%%%%%%%%%%%%%%%%%%%%%%%%%%%%%
%%%%%%%%%%%%%%%%%%%%%%%%%%%%%%%%%%%%%%%%%%%%%%%%%%
\Section{Affine and {\L}ukasiewicz Logic}
\label{sec:logics-and-algebras}
%%%%%%%%%%%%%%%%%%%%%%%%%%%%%%%%%%%%%%%%%%%%%%%%%%
%%%%%%%%%%%%%%%%%%%%%%%%%%%%%%%%%%%%%%%%%%%%%%%%%%

%%%%%%%%%%%%%%%%%%%%%%%%%%%%%%%%%%%%%%%%%%%%%%%%%%
\subsection{Language}
%%%%%%%%%%%%%%%%%%%%%%%%%%%%%%%%%%%%%%%%%%%%%%%%%%
We work in a language, $\cal L$, built from a countable set of propositional variables
$\Var = \{P_1, P_2, \ldots\}$, the constant $\F$ (falsehood) and the binary
connectives $\Lolly$ (implication) and $\iAnd$ (conjunction).
We write $\Lnot{A}$ for $A \Lolly~\F$ and $\T$ for $\F~\Lolly~\F$.
Our choice of notation for connectives is that commonly used for
affine logic, since all the systems
we consider will be extensions of intuitionistic affine logic.
% Our use of $\F$ rather than $\T$ for falsehood is taken from continuous logic
% \cite{ben-yaacov-pedersen09}, which motivated our work in this area.
% In keeping with this convention, we will order propositions by increasing logical strength, so that $A \Whence{} B$ means $A$ is at least as strong as $B$, i.e. that $A \Lolly B$ is provable.

\begin{figure}[t]
\begin{center}
\begin{tabular}{|cc|}\hline
& \\
%%
%
% Lolly introduction
%
\begin{prooftree}
\Gamma, A \vdash B
\justifies
\Gamma \vdash A \Lolly B
\using{\LI}
\end{prooftree} &
%
% Lolly elimination
%
\begin{prooftree}
\Gamma \vdash A \quad \Delta \vdash A \Lolly B
\justifies
\Gamma, \Delta \vdash B
\using{\LE}
\end{prooftree} \\\ &  \\
%%
%
% Conjunction introduction
%
\; \begin{prooftree}
\Gamma \vdash A \quad \Delta \vdash B
\justifies
\Gamma, \Delta \vdash A \iAnd B
\using{\CI}
\end{prooftree}
&
%
% Conjunction elimination
%
\begin{prooftree}
\Gamma \vdash A \iAnd B \quad \Delta, A, B \vdash C
\justifies
\Gamma, \Delta \vdash C
\using{\CE}
\end{prooftree} \\[6mm]
\hline
\end{tabular}
\caption{Natural deduction (in sequent-style) rules for $\iAnd$ and $\Lolly$}
\label{fig:sequent-rules}
\end{center}
\end{figure}

As usual, we adopt the convention that $\Lolly$ associates to the right and has
lower precedence than $\iAnd$, which in turn has lower precedence than $\Lnot{(\cdot)}$.
So, for example, the brackets in $(A \iAnd  (\Lnot{B})) \Lolly (C \Lolly (D \iAnd
F))$ are all redundant, while those in $\Lnot{(((A \Lolly B) \Lolly C) \iAnd D)}$
are all required.

%%%%%%%%%%%%%%%%%%%%%%%%%%%%%%%%%%%%%%%%%%%%%%%%%%
\subsection{The logics}
%%%%%%%%%%%%%%%%%%%%%%%%%%%%%%%%%%%%%%%%%%%%%%%%%%

In this section we give \emph{natural deduction} systems (in sequent style) for the logics we will study.
The judgments of the calculi are sequents $\Gamma \vdash A$ where the {\em
context} $\Gamma$ is a multiset of formulas and $A$ is a formula.
The rules of inference for \emph{all} the calculi
comprise the sequent formulation of a natural
deduction system shown in Figure~\ref{fig:sequent-rules}.

The six calculi are defined by adding to the rules of Figure~\ref{fig:sequent-rules} some or all of the following axiom schemata:
\emph{assumption $\ASM$, contraction $\CON$}, {\it ex falso quodlibet} $\EFQ$,
\emph{double negation elimination} $\DNE$, and \emph{commutativity of weak conjunction} $\CWC$,
defined in Figure \ref{fig:sequent-axioms}. 
% As we will discuss in Section \ref{sec-derived}, we can think of
% $A \iAnd (A \Lolly B)$ as a form of conjunction which, in the absence of contraction, is weaker than $A \iAnd B$, in general.
% Given $\ASM$ and the rules for $\iAnd$, we can view $\CWC$ as asserting that this weak form of conjunction is commutative.
%
The six calculi and their axiom schemata are as defined in Table~\ref{six-calculi}.

The systems $\ALi$, $\ALc$, $\LLi$ and $\LLc$ are intuitionistic and classical variants
of affine logic and {\L}ukasiewicz logic.
$\IL$ and $\BL$ as we shall see shortly are the usual intuitionistic and classical logic.
The relationship between the six logics is depicted in Figure~\ref{fig:logics}.

As our axiom schemata all allow additional premisses $\Gamma$ in the context, the following rule of weakening 
\[
\begin{prooftree}
\Gamma \vdash B
\justifies
\Gamma, A \vdash B
\using {\WK}
\end{prooftree}
\]
is admissible in all our logics, since given a proof tree with
$\Gamma \vdash B$ at the root, we may obtain a proof of $\Gamma, A \vdash B$
by adding $A$ to the context of every sequent on \emph{some} path from the root to a leaf (axiom). Also, note that in intuitionistic affine logic $\ALi$, and hence
in all the logics,  the contraction axiom $\CON$ is inter-derivable with the contraction rule
\[
\begin{prooftree}
\Gamma, A, A \vdash B
\justifies
\Gamma, A \vdash B
\using {\CONR}
\end{prooftree}
\]
Thus with $\CON$ we have the structural rules of weakening and contraction, which
proves our claim that $\IL$ and $\BL$ are the usual intuitionistic and classical propositional logics.

\begin{figure}[!h]
\begin{center}
\begin{tabular}{|cc|}
\hline
 & \\
\begin{prooftree}
\justifies
\Gamma, A \vdash A
\using{\ASM}
\end{prooftree}
&
\begin{prooftree}
\justifies
\Gamma, A \vdash A \iAnd A
\using{\CON}
\end{prooftree} \\
& \\
\begin{prooftree}
\justifies
\Gamma, \F~\vdash A
\using{\EFQ}
\end{prooftree} &
\begin{prooftree}
\justifies
\Gamma, \neg \neg A \vdash A
\using{\DNE}
\end{prooftree} \\ 
& \\
\multicolumn{2}{|c|}{%
\begin{prooftree}
\justifies
\Gamma, A, A \Lolly B \vdash B \iAnd (B \Lolly A)
\using{\CWC}
\end{prooftree} } \\ 
& \\
\hline
\end{tabular}
\caption{Sequent Calculus Axioms}
\label{fig:sequent-axioms}
\end{center}
\end{figure}

\begin{table}
\begin{center}
\begin{tabular}{|c|l|} \hline
Calculus & Axiom Schemata \\ \hline
$\ALi$ & $\ASM$, $\EFQ$\\ \hline
$\ALc$ & $\ASM$, $\EFQ$, $\DNE$ \\ \hline
$\LLi$ & $\ASM$, $\EFQ$, $\CWC$, \\ \hline
$\LLc$ & $\ASM$, $\EFQ$, $\CWC$, $\DNE$ \\ \hline
$\IL$  & $\ASM$, $\EFQ$, $\CON$, \\ \hline
$\BL$  & $\ASM$, $\EFQ$, $\CON$, $\DNE$ \\ \hline
\end{tabular}
\end{center}
\caption{The Six Calculi}
\label{six-calculi}
\end{table}

%\begin{Theorem} \label{thm-ml-il} $\IL$ and $\BL$ are  the implication-conjunction fragments of the usual intuitionistic and classical logics.
%\Done
%\end{Theorem}

Many of the results in this paper involve the derivability of a particular sequent in one of our calculi above (mainly $\LLi$). When deriving these, we will make clear in the statement of the result which logic we are using, and will present proofs as sequences of formulas, all of which are either an assumption, an axiom, or a consequence of previously derived formulas. We illustrate this with the following basic result:

\begin{Lemma}[$\LLi$] \label{cwc-lub} If $C \proves A$ and $C \proves B$ then $C \proves A \iAnd (A \Lolly B)$.
\end{Lemma}
\Proof \\[-6mm]
\begin{align}
& C \Lolly A 					& \using{Given} \label{cwc-lub-Given1} \\
& C \Lolly B 					& \using{Given} \label{cwc-lub-Given2} \\
& C 							& \using{Given} \label{cwc-lub-Given3} \\
& A \iAnd (A \Lolly C)				& \using{By (\ref{cwc-lub-Given1}), (\ref{cwc-lub-Given3}) and \eCWC} \label{cwc-lub-Dev2} \\
& A \iAnd (A \Lolly B)	\hspace{45mm} & \using{By (\ref{cwc-lub-Given2}) and (\ref{cwc-lub-Dev2})} \tag*{\Done}
\end{align}

The above lemma shows that (over $\LLi$) $A \iAnd (A \Lolly B)$ is the weakest formulas that is stronger than both $A$ and $B$. Note that since we do not have contraction in $\LLi$, in the above proof it is important that (\ref{cwc-lub-Given3}) is only used once, and that its consequence (\ref{cwc-lub-Dev2}) is also only used once, and so on. \\[2mm]
{\bf Notation}. We will apply the $\CWC$ axiom in slightly different ways. For instance, in the proof above we had derived $C$ and $C \Lolly B$, and by $\CWC$ were able to conclude $B \iAnd (B \Lolly C)$. In some cases we will find it more convenient to state the two conclusions $B$ and $B \Lolly C$ in separate lines of the proof, especially when these are then used in different ways later on (e.g. see proof of Lemma \ref{lemma156}). We write the justification for $B \Lolly C$ as ``$\using{Ditto}$''.  \\[-2mm]

The rules of Figure~\ref{fig:sequent-rules}
and the axioms of Figure~\ref{fig:sequent-axioms}
are closed under substitution of formulas for variables.
Hence a substitution instance of a theorem in any of our logics is again
a theorem of that logic. When reading a result such as
Lemma~\ref{cwc-lub}, it is immaterial whether one views the letters $A$,
$B$ and $C$ as metavariables ranging over $\lL$ or as specific variables in
$\Var \subset \lL$.

As mentioned in the introduction, in \cite{Ciabattoni:1997} one can find an earlier proposal of viewing {\L}ukasiewicz logic as an extension of affine logic. In that context, a rule called $(\&, l_c)$ was added to classical affine logic in order to obtain classical {\L}ukasiewicz logic. It is easy to check that in the presence of weakening the premise of that rule is derivable, which means that over affine logic the rule $(\&, l_c)$ should be viewed as the axiom schema $A \mathop{\&} B \vdash A \iAnd (A \Lolly B)$. Let us assume, for the sake of argument, that we extend $\LLi$ with the additive conjunction $A \mathop{\&} B$, with the usual rules, as in \cite{Ciabattoni:1997}. By Lemma \ref{cwc-lub} above, and since $A \mathop{\&} B \vdash A$ and $A \mathop{\&} B \vdash B$, we have that over $\ALi$ the axiom $(\&, l_c)$ follows from $\CWC$. Conversely, since $A \mathop{\&} B$ is commutative, and $A \iAnd (A \Lolly B) \vdash A \mathop{\&} B$, it follows that over $\ALi$ the axiom $\CWC$ follows from $(\&, l_c)$. Therefore, one can indeed obtain $\LLi$ by adjoining either of the axioms to $\ALi$. 
Our choice here is to work with $\CWC$, and have $A\mathop{\&}B$ as a derived connective, as we will see in the next section. 
% However, by working with $\CWC$ we could restrict ourselves to the fragment of affine logic without the additive conjunction $A \mathop{\&} B$, but only containing implication $(\Lolly)$ and the multiplicative conjunction $(\iAnd)$. 

%\begin{table}
%\[
%\begin{array}{|c|l|} \hline
%\axComp& (A \Lolly B) \Lolly (B \Lolly C) \Lolly (A \Lolly C) \\[0.5mm] \hline
%\axComm& A \iAnd B \Lolly B \iAnd A \\[0.5mm] \hline
%\axCurry& (A \iAnd  B \Lolly C) \Lolly (A \Lolly B \Lolly C) \\[0.5mm] \hline
%\axUncurry& (A \Lolly B \Lolly C) \Lolly  (A \iAnd B \Lolly C) \\[0.5mm] \hline
%\axWk& A \iAnd B \Lolly A \\[0.5mm] \hline
%\axEFQ& \F \Lolly A \\[0.5mm] \hline
%\axDNE& \Lnot{\Lnot{A}} \Lolly A \\[0.5mm] \hline
%\axCWC& A \iAnd (A \Lolly B) \Lolly B \iAnd (B \Lolly A) \\[0.5mm] \hline
%\axCon& A \Lolly A \iAnd A \\[0.5mm] \hline
%\end{array}
%\]
%\caption{Hilbert-style Axiom Schemata}
%\label{table-hilbert-systems}
%\end{table}
%
Particularly in the literature on {\L}ukasiewicz logic the
systems that we have presented above in natural (in sequent style) deduction are traditionally
presented as Hilbert-style systems with {\it modus ponens} as the
only rule of inference (see \cite{Hajek98}, Def. 3.1.3, for a Hilbert-style presentation of $\LLc$). 
%Table \ref{table-hilbert-systems} defines the axiom
%schemata {\em composition, commutativity of conjunction,
%currying, uncurrying, weakening, {\it ex falso quodlibet},
%double negation elimination, commutativity of weak
%conjunction and contraction} that feature in these systems.
%
%We can then define six Hilbert-style axiom systems as shown in Table \ref{six-hilbert-systems}.
It can be shown that the two presentations are equivalent
in the sense that the sequent $A_1, A_2, \ldots, A_n \vdash A$ is derivable in one of logics iff $A_1 \Lolly A_2 \Lolly \ldots \Lolly A_n \Lolly A$ is
derivable in the corresponding Hilbert style system.
%
%\begin{table}
%\[
%\begin{array}{|c|l|} \hline
%\Hilb{\ALi} & \axComp + \axComm + \axCurry + \axUncurry + \axWk +\axEFQ \\[0.5mm] \hline
%\Hilb{\LLi} & \Hilb{\ALi} + \axCWC \\[0.5mm] \hline
%\Hilb{\IL}  & \Hilb{\ALi} + \axCon \\[0.5mm] \hline
%\Hilb{\ALc} & \Hilb{\ALi} + \axDNE \\[0.5mm] \hline
%\Hilb{\LLc} & \Hilb{\LLi} + \axDNE \\[0.5mm] \hline
%\Hilb{\BL}  & \Hilb{\IL}  + \axDNE \\[0.5mm] \hline
%\end{array}
%\]
%%
%\caption{The Six Hilbert Systems}
%\label{six-hilbert-systems}
%\end{table}

%%%%%%%%%%%%%%%%%%%%%%%%%%%%%%%%%%%%%%%%%%%%%%%%%%
\Subsection{Derived connectives}
%%%%%%%%%%%%%%%%%%%%%%%%%%%%%%%%%%%%%%%%%%%%%%%%%%
\label{sec:derived}

In additon to the primitive connectives $\iAnd$ and $\Lolly$, we will make extensive use of the following four \emph{derived} binary connectives $\CCb{A}{B}$, $\CKb{A}{B}$, $\CMb{A}{B}$, $\oCKb{A}{B}$ defined as follows:
\[
\begin{array}{lcll}
	\CCb{A}{B} & \equiv & A \iAnd (A \Lolly B) & \mbox{(weak conjunction)} \\[2mm]
	\CKb{A}{B} & \equiv & (B \Lolly A) \Lolly A & \mbox{(strong disjunction)} \\[2mm]
	\CMb{A}{B} & \equiv & A \Lolly A \iAnd B & \mbox{(strong implication)} \\[2mm]
	\oCKb{A}{B} & \equiv & \Lnot{A} \iAnd (B \Lolly A) \quad \quad & \mbox{(NOR binary connective)}
\end{array}
\]
%
%Classically (i.e. over $\CL$) these are indeed equivalent to $A$ and $B$, $A$ or $B$, $A$ implies $B$, and neither $A$ nor $B$, respectively. 

Recall that we are assuming conjunction binds more strongly than the implication, so that $\CMb{A}{B}$ is $A \Lolly (A \iAnd B)$. For the new connectives we will also use the convention that $\CC, \CK$ and $\oCK$ all bind more strongly than $\CM$. So $\CMb{(\CKb{A}{B})}{(\CCb{C}{D})}$, for instance, may be written as $\CMb{\CKb{A}{B}}{\CCb{C}{D}}$.

$\ALi$ cannot prove the commutativity of ${\CCb{\relax}{\relax}}$ and ${\CKb{\relax}{\relax}}$. $\LLi$ adds the commutativity of ${\CCb{\relax}{\relax}}$ to $\ALi$ as an axiom schema, but still can't prove the commutativity of ${\CKb{\relax}{\relax}}$. In $\LLi$, the weak conjunction $\CCb{A}{B}$ behaves like the additive conjunction $A \& B$ of linear logic. $\LLc$ has been defined above as $\LLi$ extended with $\DNE$, but, as may been seen by simplifying $\CKb{\bot}{A}$ and $\CKb{A}{\bot}$, one also obtains $\LLc$ from $\LLi$ by adding the commutativity of ${\CKb{\relax}{\relax}}$ as an axiom schema. In $\LLc$, the strong disjunction $\CKb{A}{B}$ then behaves like the additive disjunction $A \oplus B$ of linear logic. In $\IL$, when full contraction is available, the two conjunctions $\CCb{A}{B}$ and $A \iAnd B$ become equivalent. However, $\CKb{\bot}{A} \equiv \Lnot{\Lnot{A}}$ while $\CKb{A}{\bot} \equiv A$, so that the commutativity of $\CK$ is intuitionistically unacceptable as it implies $\DNE$.

% Note that over $\ALi$, ${\CCb{\relax}{\relax}}$ and ${\CKb{\relax}{\relax}}$ are not provably \emph{commutative} operations. In fact, $\LLi$ is obtained from $\ALi$ by adding the commutative of ${\CCb{\relax}{\relax}}$ as an axiom schema, which makes ${\CCb{\relax}{\relax}}$ have the same properties as the additive conjunction of linear logic $\&$. 

% We justify our notation by observing that when $\iAnd$ and $\Lolly$ are replaced by the standard conjunction and implication of \emph{classical logic} then $\CCb{A}{B}, \CKb{A}{B}, \CMb{A}{B}$ are indeed equivalent to the standard conjunction, disjunction and implication. Moreover, $\oCKb{A}{B}$ is equivalent to the NOR binary connective. In affine logic, however, it is not even possible to prove that  $\CCb{A}{B}$ and $\CKb{A}{B}$ are \emph{commutative}.
%The commutativity of $\CCb{A}{B}$ is the axiom schema $\CWC$ that defines $\LLi$, while the commutativity of $\CKb{A}{B}$ is axiom (A3) of Remark \ref{Remark-rose}.

We have chosen our notation so that in each of the derived connectives the left operand appears both positively and negatively while the right operand appears only positively in $\CC, \CK$ and $\CM$ and only negatively in $\oCK$.

\begin{Definition} Let $T$ be any of the extensions of $\ALi$ discussed above. We write $A \Equiv{T} B$ if $A \proves B$ and $B \proves A$ in $T$. When the $T$ in question is clear from the context we just write $A \Equiv{} B$.
\end{Definition}

Let $A[B]$ be a formula that contains $B$ as a subformula. It is easy to show, by induction on $A$, that if $B \Equiv{} C$ then $A[B] \Equiv{} A[C]$.

We conclude this section with a short list of basic theorems of $\ALi$ which will prove very useful in the sequel. 
%Recall that $A \Whence{} B$ stands for $A \vdash B$; and that $A \Equiv{} B$ stands for $A \vdash B$ and $B \vdash A$.

\begin{Lemma}[$\ALi$] \label{ali-basic-lemma} The following have simple and short derivations:
\begin{description}
	\item[(i)] $A \proves \CKb{B}{A}$, in particular, taking $B = \F$, $A \proves \Lnot{\Lnot{A}}$
	\item[(ii)] $A \proves \CMb{B}{A}$
	\item[(iii)] $A \iAnd B \Equiv{} A \iAnd (\CMb{A}{B})$
	\item[(iv)] $\CKb{C}{(A \Lolly B)} \proves (\CKb{C}{A}) \Lolly (\CKb{C}{B})$
	\item[(v)] $\Lnot{\Lnot{(\Lnot{A} \Lolly \Lnot{B})}} \Equiv{} \Lnot{A} \Lolly \Lnot{B}$
	\item[(vi)] $\Lnot{\Lnot{A}} \Lolly \Lnot{\Lnot{B}} \Equiv{} A \Lolly \Lnot{\Lnot{B}}$
	\item[(vii)] $\Lnot{(\Lnot{\Lnot{A}} \iAnd \Lnot{\Lnot{B}})} \Equiv{} \Lnot{(A \iAnd B)}$
	\item[(viii)] $\Lnot{\Lnot{(A \iAnd B)}} \Equiv{} \Lnot{(\Lnot{\Lnot{A}} \Lolly \Lnot{B})}$
\end{description}
\end{Lemma}
\Proof Easy, making much use of $\Lnot{\Lnot{\Lnot{A}}} \Equiv{} \Lnot{A}$ and
$A \Lolly \Lnot{B} \Equiv{} \Lnot{(A \iAnd B)} \Equiv{} B \Lolly \Lnot{A}$. For instance, we can prove $(vi)$ via the following chain of simple equivalences:
\[
\begin{array}{c@{}c@{}c@{}c@{}c@{}c@{}c@{}cr}
& \Lnot{\Lnot{A}} \Lolly \Lnot{\Lnot{B}} 
	& {}\Equiv{} & \Lnot{(\Lnot{\Lnot{A}} \iAnd \Lnot{B})}
	& {}\Equiv{} & \Lnot{(\Lnot{B} \iAnd \Lnot{\Lnot{A}})}
	& {}\Equiv{} & \Lnot{B} \Lolly \Lnot{\Lnot{\Lnot{A}}} & \\[1mm]
{}\Equiv{} & \Lnot{B} \Lolly \Lnot{A} & \Equiv{} & \Lnot{(\Lnot{B} \iAnd A)} 
	& \Equiv{} & \Lnot{(A \iAnd \Lnot{B})}  & \Equiv{} & A \Lolly \Lnot{\Lnot{B}}  & \Done
\end{array}
\]

%%%%%%%%%%%%%%%%%%%%%%%%%%%%%%%%%%%%%%%%%%%%%%%%%%
%%%%%%%%%%%%%%%%%%%%%%%%%%%%%%%%%%%%%%%%%%%%%%%%%%
\Section{Negative Translations of Affine Logic}
%%%%%%%%%%%%%%%%%%%%%%%%%%%%%%%%%%%%%%%%%%%%%%%%%%
%%%%%%%%%%%%%%%%%%%%%%%%%%%%%%%%%%%%%%%%%%%%%%%%%%
\label{sec:neg-trans}

In \cite{Troelstra(73)}, Troelstra identifies certain requirements on a translation of classical logic into  intuitionistic logic and shows that any two translations satisfying these requirements are intuitionistically equivalent. To set up the analogue of this characterisation in our substructural setting, we first define the notion of {\em negative formula} in the language $\lL$.

\begin{Definition} The set $\Negative$ of \emph{negative formulas} is defined inductively as
\begin{itemize}
	\item $\F~\in \Negative$,
	\item if $A \in \Negative$ and $B \in \Negative$ then $A \iAnd B \in \Negative$,
	\item if $B \in \Negative$ then  $A \Lolly B \in \Negative$.
\end{itemize}
\end{Definition}

We can now formulate an adaptation of Troelstra's requirements:

\begin{Definition} Let ${\bf L}$ be a fragment of intuitionistic logic $\IL$ over the language ${\cal L}$. 
A formula translation $(\cdot)^\dagger \colon {\cal L} \to {\cal L}$ is called a \emph{negative translation for {\bf L}} if the following holds for every formula $A$ in the language of ${\bf L}$ 
\begin{description}
	\item[{\bf (NT1)}] ${\bf L}$ proves $A^\dagger \vdash B$ and $B \vdash A^\dagger$, for some $B \in \Negative$.
	\item[{\bf (NT2)}] ${\bf L} + \DNE$ proves $A^\dagger \vdash A$ and $A \vdash A^\dagger$.
	\item[{\bf (NT3)}] if ${\bf L} + \DNE$ proves $\vdash A$ then ${\bf L}$ proves $\vdash A^\dagger$.
\end{description}
\end{Definition}

Conditions {\bf (NT1)}, {\bf (NT2)} and {\bf (NT3)} correspond to Troelstra's \cite[Section 10]{Troelstra(73)} ($iii$), ($i$) and ($ii$), respectively. We have rearranged them as we will show that in $\LLi$ condition  {\bf (NT3)} is redundant (Theorem \ref{theorem-main}). It is often the case in practice that $A^\dagger \in \Negative$, so that {\bf (NT1)} holds trivially.

In this section we shall consider the four well-known negative translations for $\IL$, namely, Kolmogorov, G\"odel, Gentzen and Glivenko, in the context of affine logic $\ALi$ (cf. \cite{FO(2012B)} for an analysis of the relationship between these translations in the setting of intuitionistic first-order logic). We prove that both Kolmogorov and G\"odel are negative translation for $\ALi$, and give counter-examples to show that Gentzen and Glivenko fail to satisfy {\bf (NT3)}. In Section \ref{sec:embedding}, however, we will see that in $\LLi$ all these formula translations are negative translations, and in fact, we will also be able to show all negative translations are provably equivalent in $\LLi$.

%%%%%%%%%%%%%%%%%%%%%%%%%%%%%%%%%%%%%%%%%%%%%%%%%%
\Subsection{Kolmogorov and G\"odel translations}
%%%%%%%%%%%%%%%%%%%%%%%%%%%%%%%%%%%%%%%%%%%%%%%%%%

First of all, we show that both the Kolmogorov and the G\"odel translations are in fact negative translations for \emph{affine} logic, i.e. no form of contraction is necessary to prove {\bf (NT1)} -- {\bf (NT3)}. Let ${\cal L}$ be the language of the theories $\ALc$ and $\ALi$.

\begin{Definition}[Kolmogorov translation \cite{Kolmogorov(25)}] For each formula $A \in {\cal L}$ associate a formula $\kTrans{A} \in {\cal L}$ inductively as follows:
\[
\begin{array}{rcl}
	\kTrans{P} & \equiv & \Lnot{\Lnot{P}} \quad\quad (\mbox{$P$ atomic}) \\[2mm]
	\kTrans{\F} & \equiv & \F \\[2mm]
	\kTrans{(A \iAnd B)} & \equiv & \Lnot{\Lnot{(\kTrans{A} \iAnd \kTrans{B})}} \\[2mm]
	\kTrans{(A \Lolly B)} & \equiv & \Lnot{\Lnot{(\kTrans{A} \Lolly \kTrans{B})}}.
\end{array}
\]
\end{Definition}

We will also consider the following negative translation which can be distilled from \cite{Goedel(33)}. In G\"odel's presentation an implication $A \Lolly B$ is translated as $\Lnot{(A \iAnd \Lnot{B})}$. We use here that in $\ALi$ this is equivalent to $A \Lolly \Lnot{\Lnot{B}}$. The translation  often referred to as the  G\"odel-Gentzen translation will be treated in the following section, where we attribute it to Gentzen. It will become clear that in the substructural setting the G\"odel translation is not the same as the Gentzen one.

\begin{Definition}[G\"odel translation \cite{Goedel(33)}] For each formula $A \in {\cal L}$ we first associate a formula $\godTransPre{A} \in {\cal L}$ inductively as follows:
\[
\begin{array}{rcl}
	\godTransPre{P} & \equiv & P \quad\quad (\mbox{$P$ atomic}) \\[2mm]
	\godTransPre{\F} & \equiv & \F \\[2mm]
	\godTransPre{(A \iAnd B)} & \equiv & \godTransPre{A} \iAnd \godTransPre{B} \\[2mm]
	\godTransPre{(A \Lolly B)} & \equiv & \godTransPre{A} \Lolly \Lnot{\Lnot{\godTransPre{B}}}.
\end{array}
\]
\end{Definition}
Then we define $\godTrans{A} = \Lnot{\Lnot{\godTransPre{A}}}$. G\"odel \cite{Goedel(33)} in fact does not need this final double negation since in Heyting arithmetic one can already prove $\Lnot{\Lnot{\godTransPre{A}}} \vdash \godTransPre{A}$. Hence in that context we can even take $\godTrans{A} = \godTransPre{A}$. In $\ALi$, however, we need the outermost double negation to make the proof of the following theorem go through.

% Recalling that $\Lnot{A} = A \Lolly~\F$, one may check that $\kTrans{(\Lnot{A})} \Equiv{} \Lnot{(\kTrans{A})}$ and $\Lnot{(\godTrans{A})} \Equiv{} \Lnot{(\godTrans{A})}$.

\begin{Theorem} \label{thm-kolm} Both the Kolmogorov translation $\kTrans{(\cdot)}$ and the G\"odel translation $\godTrans{(\cdot)}$ are negative translations for $\ALi$.
\end{Theorem}
\Proof In the case of Kolmogov we have: \\
{\bf (NT1)}.~Trivial since $\kTrans{A} \in \Negative$. \\[1mm]
{\bf (NT2)}.~Clearly $\ALc = \ALi + \DNE$ proves $A \Equiv{} \kTrans{A}$. \\[1mm]
{\bf (NT3)}.~Finally, by using Lemma \ref{ali-basic-lemma}, it is easy to show that if $\Gamma \vdash A$ is provable in $\ALc$ then $\kTrans{\Gamma} \vdash \kTrans{A}$ is provable in $\ALi$, where $\kTrans{\Gamma}$ abbreviates $\kTrans{B_0}, \ldots \kTrans{B_n}$. This can be shown by induction on the derivation of the sequent $\Gamma \vdash A$. \\[1mm]
Exactly the same approach works for the G\"odel translation. \Done

\Subsection{Gentzen and Glivenko translations}
%%%%%%%%%%%%%%%%%%%%%%%%%%%%%%%%%%%%%%%%%%%%%%%%%%
\label{gentzen-fail}

For both the Gentzen and the Glivenko translations (defined below) a corresponding Theorem \ref{thm-kolm} no longer holds for $\ALi$. These translations rely on uses of contraction which are not available in affine logic. Nevertheless, we will find that the amount of contraction available in {\L}ukasiewicz logic, via $\CWC$, is sufficient for these translations to go through (Section \ref{sec:embedding}). 

The Gentzen negative translation works by adding double negations on all the atoms of a given formula:

\begin{Definition}[Gentzen translation \cite{Gentzen(33)}] \label{def:gentzen} For each formula $A \in {\cal L}$ associate a formula $\ggTrans{A} \in {\cal L}$ inductively as follows:
\[
\begin{array}{rcl}
	\ggTrans{P} & \equiv & \Lnot{\Lnot{P}} \quad\quad (\mbox{$P$ atomic}) \\[2mm]
	\ggTrans{\F} & \equiv & \F \\[2mm]
	\ggTrans{(A \iAnd B)} & \equiv & \ggTrans{A} \iAnd \ggTrans{B} \\[2mm]
	\ggTrans{(A \Lolly B)} & \equiv & \ggTrans{A} \Lolly \ggTrans{B}.
\end{array}
\]
As $\Lnot{A} = A \Lolly~\F$, we have that $\ggTrans{(\Lnot{A})}$ is equivalent to $\Lnot{\ggTrans{A}}$. 
\end{Definition}

The Glivenko negative translation simply doubly negates the whole formula:

\begin{Definition}[Glivenko translation \cite{Glivenko(29)}] \label{def:glivenko} Given a formula $A \in {\cal L}$ define its Glivenko translation $\glTrans{A}$ as $\glTrans{A} \equiv \Lnot{\Lnot{A}}$.
\end{Definition}

We have shown in \cite[Theorems 3.9 and 3.10]{arthan-oliva-2021} that neither translation is a negative translation for $\ALi$, but they do not fail simultaneously, i.e. there are extensions of $\ALi$ such that the $\glTrans{(\cdot)}$ translation works but $\ggTrans{(\cdot)}$ translation does not, and vice-versa. The proofs make use of the following two classically provable formulas which are unprovable in $\ALi$: $A \equiv \Lnot{\Lnot{P}} \Lolly P$ and $B \equiv \Lnot{\Lnot{(P \iAnd Q)}} \Lolly P \iAnd Q$, and shows that $\ALi$ proves neither $\glTrans{A}$ nor $\ggTrans{B}$.

%%%%%%%%%%%%%%%%%%%%%%%%%%%%%%%%%%%%%%%%%%%%%%%%%%
%%%%%%%%%%%%%%%%%%%%%%%%%%%%%%%%%%%%%%%%%%%%%%%%%%
\Section{Homomorphism Properties of Double Negation in $\LLi$}
\label{sec:lli}
%%%%%%%%%%%%%%%%%%%%%%%%%%%%%%%%%%%%%%%%%%%%%%%%%%
%%%%%%%%%%%%%%%%%%%%%%%%%%%%%%%%%%%%%%%%%%%%%%%%%%

Our goal in this section is to find $\LLi$ derivations of some important theorems about the primitive and derived connectives. These include:
% basic properties of $\CK, \CM$ and $\CC$ (Section \ref{sec:basic-ili-k-m-c});
\begin{itemize}
	\item a derivation of $\Lnot{\Lnot{\DNE}}$  (Corollary \ref{thm139}); 
	\item a duality property between $\CK$ and $\oCK$ (Theorem \ref{k-negated}); and,
	\item homomorphism properties of the double negation mapping $A \mapsto \Lnot{\Lnot{A}}$ with respect to both implication (Section \ref{homo-dn-lolly}) and conjunction (Section \ref{homo-dn-tensor}); 
%	\item a collection of ``De Morgan" properties (Section \ref{sec-de-morgan}).
\end{itemize}

As we have already remarked, the derivations we will give have been extracted by analysis of computer-generated proofs found by the Prover9 automated theorem-prover.
Our contribution was to propose conjectures to Prover9, to study the
machine-oriented proofs it found and to present the proofs in a
human-intelligible form by breaking them down into structurally interesting
lemmas. This was an iterative process since often Prover9 was able to find
simpler proofs of a lemma when presented with it as a conjecture in isolation.
As discussed in \cite{arthan-oliva-2021}, in cases when Prover9 was unable to find a proof, Mace4 was often able to find
a counter-model: a finite model of the logic in question in which the
conjecture can be seen to fail.

It follows from work on commutative GBL-algebras that $\LLi$ is decidable \cite{Bova:2009}. However the problem is PSPACE-complete. 

% In~\cite{arthan-oliva14b}, we give a simple indirect method for demonstrating that a formula is valid in intuitionistic {\L}ukasiewicz Logic, a heuristic method which we have used extensively in parallel with attempts to find explicit proofs and counter-examples with Prover9 and Mace4.

%%%%%%%%%%%%%%%%%%%%%%%%%%%%%%%%%%%%%%%%%%%%%%%%%%
\Subsection{Basic identities on $\CK, \CM, \CC$}
%%%%%%%%%%%%%%%%%%%%%%%%%%%%%%%%%%%%%%%%%%%%%%%%%%
\label{sec:basic-ili-k-m-c}

We start by proving in $\LLi$ several useful results about the derived connectives $\CK, \CC$ and $\CM$.
% The main identity is $A \iAnd B \Equiv{} A \iAnd (\CKb{B}{(\CMb{A}{B})})$ which will be extensively used throughout the rest of the paper. The left-to-right implication is obvious, as $B \Whence{} \CKb{B}{X}$. The converse, however, says that in the context $A \iAnd (\cdot)$ we have $\CKb{B}{(\CMb{A}{B})} \Whence{} B$.

%\begin{Lemma}[$\LLi$] \label{k-cwc} $A \Equiv{} (\CKb{A}{B}) \iAnd (B \Lolly A)$
%\end{Lemma}
%%
%\Proof We have
%%
%\begin{align*}
%A
%	& \Equiv{} A \iAnd \underline{(A \Lolly (B \Lolly A))} \by{\WK} \\[0mm]
%	& \Equiv{} ((B \Lolly A) \Lolly A) \iAnd (B \Lolly A) \by{\CWC} \\[1mm]
%	& \Equiv{} (\CKb{A}{B}) \iAnd (B \Lolly A). \by{\mbox{def $\CK$}}
%\end{align*}
%%
%Recall that we underline easily proved conjuncts which are either inserted or deleted. \Done

\begin{Lemma}[$\LLi$] \label{lemma156} $\CKb{B}{(\CMb{A}{B})} \proves \CMb{A}{B}$
\end{Lemma}
\Proof Since $\CMb{A}{B} \equiv A \Lolly A \iAnd B$, we will show $A, \CKb{B}{(\CMb{A}{B})} \proves A \iAnd B$:
\begin{align}
& A 							& \using{Given} \label{Given1} \\
& \CKb{B}{(\CMb{A}{B})} \;\; (\mbox{i.e., } ((\CMb{A}{B}) \Lolly B) \Lolly B) & \using{Given} \label{Given2} \\
& A \Lolly ((\CMb{A}{B}) \Lolly B)	& \using{Derivable} \label{Dev1} \\
& (\CMb{A}{B}) \Lolly B			& \using{By (\ref{Given1}), (\ref{Dev1}) and $\eCWC$} \label{Dev2} \\
& ((\CMb{A}{B}) \Lolly B) \Lolly A	&  \using{Ditto} \label{Dev3} \\
& B 							& \using{By (\ref{Given2}), (\ref{Dev2}) and $\eLE$} \label{Dev4} \\
& B \Lolly (\CMb{A}{B})			& \using{By Lemma \ref{ali-basic-lemma} $(ii)$} \label{Dev5} \\
& \CMb{A}{B} \; (\mbox{i.e., } A \Lolly A \iAnd B) & \using{By (\ref{Dev4}), (\ref{Dev5}) and $\eCWC$} \label{Dev6} \\
& (\CMb{A}{B}) \Lolly B			&  \using{Ditto} \label{Dev7} \\
& A 							& \using{By (\ref{Dev3}), (\ref{Dev7}) and $\eLE$} \label{Dev8} \\
& A \iAnd B \hspace{50mm}		& \using{By (\ref{Dev6}), (\ref{Dev8}) and $\eLE$} \tag*{\Done}
\end{align}

%Assume $(i)$ $A$ and $(ii)$ $(\CKb{B}{(\CMb{A}{B})})$, i.e. $((\CMb{A}{B}) \Lolly B) \Lolly B$, and note that $(iii)$ $(A \Lolly ((\CMb{A}{B}) \Lolly B))$
%is provable. From $(i)$ and $(iii)$ and $\CWC$ we obtain $(iv)$ $(\CMb{A}{B}) \Lolly B$ and $(v)$ $((\CMb{A}{B}) \Lolly B)) \Lolly A$. From $(ii)$ and $(iv)$, by MP, we get $(vi)$ $B$. By $(vi)$ and $B \Lolly (\CMb{A}{B})$ (Lemma \ref{ali-basic-lemma} $(ii)$), using $\CWC$, we have $(vii)$ $\CMb{A}{B}$, i.e. $A \Lolly A \iAnd B$, and $(viii)$ $(\CMb{A}{B}) \Lolly B$. By $(v)$ and $(viii)$, by MP, we have $A$, which by $(vii)$ gives us $A \iAnd B$. \Done \\

The following lemma is used in Section \ref{homo-dn-tensor}. It shows that from $A \Lolly C$ and $C \Lolly B$ we can conclude more than $A \Lolly C$.

\begin{Lemma}[$\LLi$] \label{guess} $A \Lolly C, C \Lolly B \proves (A \Lolly B) \iAnd (\CCb{A}{B} \Lolly C)$
\end{Lemma}
\Proof \\[-6mm]
\begin{align}
& A \Lolly C 					& \using{Given} \label{guess-Given1} \\
& C \Lolly B 					& \using{Given} \label{guess-Given2} \\
& (A \Lolly C) \Lolly (A \Lolly B)		& \using{From (\ref{guess-Given2}), easy} \label{guess-Dev1} \\
& (A \Lolly B) \iAnd ((A \Lolly B) \Lolly (A \Lolly C)) & \using{By (\ref{guess-Given1}) and (\ref{guess-Dev1}) and \eCWC} \label{guess-Dev2} \\
& (A \Lolly B) \iAnd (\CCb{A}{B} \Lolly C)	& \using{By (\ref{guess-Dev2}) and Def. of $\CCb{A}{B}$} \tag*{\Done}
\end{align}

So far, we have not used the constant $\F$. We now prove a few basic properties of $\Lnot{(\cdot)}$ and its relation to the derived connectives. 
% or negation (recall that we define negation by $\Lnot{A} = A \Lolly \F$).

\begin{Lemma}[$\LLi$] \label{lemma165} $\CKb{B}{A} \vdash \CMb{\Lnot{A}}{B}$
\end{Lemma}
\Proof \\[-6mm]
\begin{align}
& \CKb{B}{A} 					& \using{Given} \label{lemma165-Given2} \\
& A \Lolly (\CMb{\Lnot{A}}{B})		& \using{Derivable, easy} \label{lemma165-Dev1} \\
& \CKb{B}{(\CMb{\Lnot{A}}{B})} 		& \using{By (\ref{lemma165-Given2}) and (\ref{lemma165-Dev1})} \label{lemma165-Dev2} \\
& \CMb{\Lnot{A}}{B} \hspace{40mm} 	& \using{By (\ref{lemma165-Dev2}) and Lemma \ref{lemma156}} \tag*{\Done}
\end{align}

It turns out that many intuitionistically unacceptable equivalences become provable in $\LLi$ ``under" a negation.
For example, our first important result is that in $\LLi$ strong implication $\CM$ is a dual of a weak conjunction $\CC$ in the sense that $\Lnot{(\CCb{A}{B})} \Equiv{} \CMb{A}{\Lnot{B}}$. This is akin to the relation between conjunction and implication $\Lnot{(A \iAnd B)} \Equiv{} A \Lolly \Lnot{B}$ which one obtains in $\ALi$ simply by currying and uncurrying. 

\begin{Theorem}[$\LLi$] \label{c-demorgan} $\Lnot{(\CCb{A}{B})} \Equiv{} \CMb{A}{\Lnot{B}}$
\end{Theorem}
\Proof The derivation of $\CMb{A}{\Lnot{B}} \vdash \Lnot{(\CCb{A}{B})}$ is straightforward. We present the derivation of the converse $\Lnot{(\CCb{A}{B})} \vdash \CMb{A}{\Lnot{B}}$,
\begin{align}
& \Lnot{(\CCb{A}{B})} \; (\mbox{i.e., } \Lnot{(A \iAnd (A \Lolly B))}) & \using{Given} \label{c-demorgan-Given1} \\
& A \Lolly (B \Lolly A)				& \using{Derivable, easy} \label{c-demorgan-Given2} \\
& \Lnot{B} \Lolly (\CKb{A}{B}) \Lolly A \iAnd \Lnot{B} & \using{Derivable, Lemma \ref{lemma165}} \label{c-demorgan-Dev0} \\
& \Lnot{(B \iAnd (B \Lolly A))}		& \using{From (\ref{c-demorgan-Given1}) and $\eCWC$} \label{c-demorgan-Dev1} \\
& (B \Lolly A) \Lolly \Lnot{B}			& \using{From (\ref{c-demorgan-Dev1}), easy} \label{c-demorgan-Dev2} \\
& (B \Lolly A) \Lolly (\CKb{A}{B}) \Lolly A \iAnd \Lnot{B} & \using{By $(\ref{c-demorgan-Dev0})$ and (\ref{c-demorgan-Dev2})} \label{c-demorgan-Dev3} \\
& A \Lolly (A \Lolly (B \Lolly A)) \Lolly A \iAnd \Lnot{B}	& \using{By (\ref{c-demorgan-Dev3}) and $\eCWC$} \label{c-demorgan-Dev4} \\
& A \Lolly A \iAnd \Lnot{B}			& \using{By (\ref{c-demorgan-Given2}) and (\ref{c-demorgan-Dev4})} \label{c-demorgan-Dev5} \\
& \CMb{A}{\Lnot{B}} 				& \using{By (\ref{c-demorgan-Dev5}) and Def. of $\CMb{A}{B}$} \tag*{\Done}
\end{align}

\begin{Corollary}[$\LLi$] \label{lemma48459} $\CMb{A}{\Lnot{B}} \Equiv{} \CMb{B}{\Lnot{A}}$
\end{Corollary}
\Proof Direct from Theorem \ref{c-demorgan}, since $\CC$ is commutative (i.e. $\CWC$). \Done

%%%%%%%%%%%%%%%%%%%%%%%%%%%%%%%%%%%%%%%%%%%%%%%%%%
\Subsection{Symmetries of $\CK$ and $\oCK$ and $\DNE$}
%%%%%%%%%%%%%%%%%%%%%%%%%%%%%%%%%%%%%%%%%%%%%%%%%%
\label{sec:sym-k-ok}

Although the commutativity of $\CK$ is clearly a classical principle, it is perhaps surprising that commutativity of $\oCKb{B}{A} $ can be proved intuitionistically.

\begin{Theorem}[$\LLi$] \label{a-lolly-b-not-a} $\oCKb{B}{A} \Equiv{} \oCKb{A}{B}$
\end{Theorem}
\Proof By symmetry it is enough to prove $\oCKb{B}{A} \proves \oCKb{A}{B}$. Recall that $\oCKb{B}{A}$ is defined as $(B \Lolly A) \iAnd \Lnot{A}$. Hence, we must show $B \Lolly A, \Lnot{A} \proves (A \Lolly B) \iAnd \Lnot{B}$, which we can do as follows:
\begin{align}
& B \Lolly A					& \using{Given} \label{a-lolly-b-not-a-Given1} \\
& \Lnot{A}						& \using{Given} \label{a-lolly-b-not-a-Given2} \\
& \Lnot{A} \Lolly (A \Lolly B)			& \using{Derivable (using \EFQ)} \label{a-lolly-b-not-a-Given3} \\
& A \Lolly B					& \using{From (\ref{a-lolly-b-not-a-Given2}), (\ref{a-lolly-b-not-a-Given3}) and $\eCWC$} \label{a-lolly-b-not-a-Dev1} \\
& (A \Lolly B) \Lolly \Lnot{A}			&  \using{Ditto}\label{a-lolly-b-not-a-Dev2} \\
& (B \Lolly A) \Lolly \Lnot{B}			& \using{From (\ref{a-lolly-b-not-a-Dev2}) and $\eCWC$} \label{a-lolly-b-not-a-Dev3} \\
& \Lnot{B} 						& \using{By (\ref{a-lolly-b-not-a-Given1}) and (\ref{a-lolly-b-not-a-Dev3})} \label{a-lolly-b-not-a-Dev4} \\
& (A \Lolly B) \iAnd \Lnot{B}	\hspace{35mm}	 & \using{By (\ref{a-lolly-b-not-a-Dev1})  and (\ref{a-lolly-b-not-a-Dev4})} \tag*{\Done}
\end{align}

A corollary of the above theorem is that the double negation of the classical axiom $\DNE$ is provable in $\LLi$:

\begin{Corollary}[$\LLi$] \label{thm139} $\Lnot{\Lnot{(\Lnot{\Lnot{A}} \Lolly A)}}$
\end{Corollary}
\Proof Let $X = (\CMb{\Lnot{A}}{A})$. Then
\begin{align}
& \Lnot{(\Lnot{\Lnot{A}} \Lolly A)}				& \using{Given} \label{thm139-Given1} \\
& \Lnot{\Lnot{A}} \Lolly (\CMb{\Lnot{A}}{A})		& \using{Derivable, easy} \label{thm139-Dev0} \\
& \Lnot{((\CMb{\Lnot{A}}{A}) \Lolly A)}			& \using{From (\ref{thm139-Given1}) using (\ref{thm139-Dev0})} \label{thm139-Dev1} \\
& A \Lolly ((\CMb{\Lnot{A}}{A}) \Lolly A) 		& \using{Derivable (using \WK)} \label{thm139-Given2} \\
& \oCKb{((\CMb{\Lnot{A}}{A}) \Lolly A)}{A}		& \using{From (\ref{thm139-Dev1}) and (\ref{thm139-Given2})} \label{thm139-Dev2} \\
& \oCKb{A}{((\CMb{\Lnot{A}}{A}) \Lolly A)}		& \using{By (\ref{thm139-Dev2}) and Theorem \ref{a-lolly-b-not-a}} \label{thm139-Dev3} \\
& (\oCKb{A}{(X \Lolly A)}) \Lolly (\Lnot{A} \iAnd (\CKb{A}{X}))	& \using{Derivable, easy} \label{thm139-Given3} \\
& \Lnot{A} \iAnd (\CKb{A}{(\CMb{\Lnot{A}}{A})})	& \using{By (\ref{thm139-Dev3}) and (\ref{thm139-Given3})} \label{thm139-Dev4} \\
& \Lnot{A} \iAnd A						& \using{By (\ref{thm139-Dev4}) and Lemma (\ref{lemma156})} \label{thm139-Dev5} \\
& \F	\hspace{55mm}	 					& \using{Easy} \tag*{\Done}
\end{align}

\begin{Remark} It is well known that the above corollary is provable in full intuitionistic logic $\IL$. The usual proof making (apparently) essential use of the full contraction axiom goes as follows. Assuming (a) $\Lnot{(\Lnot{\Lnot{A}} \Lolly A)}$ we must derive a contradiction. First use (a) to derive $\Lnot{A}$, by $\WK$. Assume also (b) $\Lnot{\Lnot{A}}$. From (b) and $\Lnot{A}$ we obtain $\F$, and hence $A$. Discharging the assumption (b) we have $\Lnot{\Lnot{A}} \Lolly A$, which by (a) gives a contradiction. Note, however, that assumption (a) was used twice. The corollary above gives us a proof using only the weak form of contraction permitted by $\CWC$.
\end{Remark}

Next we present a theorem showing that the NOR connective $\oCKb{A}{B}$ is indeed the negation of the disjunction $\CK$, a fact which holds in full intuitionistic logic $\IL$, but again, via  a simple proof that appears to make essential use of the full contraction axiom. 

\begin{Theorem}[$\LLi$] \label{k-negated} $\Lnot{(\CKb{A}{B})} \Equiv{} \oCKb{A}{B}$
\end{Theorem}
\Proof As usual one of the directions is easy, in this case $\oCKb{A}{B} \vdash \Lnot{(\CKb{A}{B})}$. We prove the other direction $\Lnot{((B \Lolly A) \Lolly A)} \proves \Lnot{A} \iAnd (B \Lolly A)$ as follows:
\begin{align}
& \Lnot{((B \Lolly A) \Lolly A)}				& \using{Given} \label{k-negated-Given1} \\
& \Lnot{((B \Lolly A) \Lolly A)} \Lolly (B \Lolly A)	& \using{Derivable} \label{k-negated-Given2} \\
& B \Lolly A							& \using{From (\ref{k-negated-Given1}) and (\ref{k-negated-Given2}) and $\eCWC$} \label{k-negated-Dev1} \\
& (B \Lolly A) \Lolly \Lnot{((B \Lolly A) \Lolly A)}	&  \label{k-negated-Dev2} \\
& \Lnot{((B \Lolly A) \iAnd ((B \Lolly A) \Lolly A))}	& \using{By (\ref{k-negated-Dev2}), easy} \label{k-negated-Dev3} \\
& \Lnot{(A \iAnd (A \Lolly (B \Lolly A)))}	& \using{By (\ref{k-negated-Dev3}) and $\eCWC$} \label{k-negated-Dev4} \\
& A \Lolly (B \Lolly A)						& \using{Derivable} \label{k-negated-Dev5} \\
& \Lnot{A}								& \using{By (\ref{k-negated-Dev4}) and (\ref{k-negated-Dev5})} \label{k-negated-Dev6} \\
& \Lnot{A} \iAnd (B \Lolly A)	\hspace{25mm}		& \using{By (\ref{k-negated-Dev1}) and (\ref{k-negated-Dev6}) } \tag*{\Done}
\end{align}

The above theorem implies the commutativity of $\CKb{A}{B}$ under a negation:

\begin{Theorem}[$\LLi$] $\Lnot{(\CKb{A}{B})} \Equiv{} \Lnot{(\CKb{B}{A})}$
\end{Theorem}
\Proof Direct from Theorems \ref{a-lolly-b-not-a} and \ref{k-negated}. \Done

%%%%%%%%%%%%%%%%%%%%%%%%%%%%%%%%%%%%%%%%%%%%%%%%%%
\subsection{Double negation homomorphism: Implication}
%%%%%%%%%%%%%%%%%%%%%%%%%%%%%%%%%%%%%%%%%%%%%%%%%%
\label{homo-dn-lolly}

We now show that (in $\LLi$) the double negation mapping $A \mapsto \Lnot{\Lnot{A}}$ is a homomorphism for implication, i.e.
\[ \Lnot{\Lnot{(A \Lolly B)}} \;\Equiv{}\; \Lnot{\Lnot{A}} \Lolly \Lnot{\Lnot{B}}. \]
We will show the same for conjunction in Section \ref{homo-dn-tensor}. Note that by definition $\CKb{\F}{~A} \equiv \Lnot{\Lnot{A}}$. Hence, it follows from Lemma \ref{ali-basic-lemma} ($iv$) that $$\Lnot{\Lnot{(A \Lolly B)}} \proves \Lnot{\Lnot{A}} \Lolly \Lnot{\Lnot{B}}$$ and hence $\Lnot{\Lnot{(A \Lolly B)}} \proves A \Lolly \Lnot{\Lnot{B}}$ is provable already in $\ALi$. We will now see that the converse implication holds in $\LLi$. Again, the fact that this holds in full intuitionistic logic is well known. See \cite{Troelstra(73)}, page 9, for instance, for an $\IL$-derivation of Theorem \ref{thm-lolly-not-not}. That derivation, however, uses the assumption $\Lnot{(A \Lolly B)}$ twice, and hence cannot be formalised in $\LLi$. 

\begin{Theorem}[$\LLi$] \label{thm-lolly-not-not} $\Lnot{\Lnot{A}} \Lolly \Lnot{\Lnot{B}} \Equiv{} \Lnot{\Lnot{(A \Lolly B)}}$
\end{Theorem}
\Proof By the remarks above, we have only the left-to-right direction to prove:
\begin{align}
& \Lnot{\Lnot{A}} \Lolly \Lnot{\Lnot{B}}				& \using{Given} \label{thm-lolly-not-not-Given1} \\
& A \Lolly \Lnot{\Lnot{A}}					& \using{Derivable} \label{thm-lolly-not-not-Given2} \\
& A \Lolly \Lnot{\Lnot{B}}					& \using{From (\ref{thm-lolly-not-not-Given1}) and (\ref{thm-lolly-not-not-Given2})} \label{thm-lolly-not-not-Dev1} \\
& (\Lnot{\Lnot{B}} \Lolly B) \Lolly (A \Lolly B)		& \using{From (\ref{thm-lolly-not-not-Dev1})} \label{thm-lolly-not-not-Dev2} \\
& \Lnot{(A \Lolly B)} \Lolly \Lnot{(\Lnot{\Lnot{B}} \Lolly B)}	& \using{From (\ref{thm-lolly-not-not-Dev2})} \label{thm-lolly-not-not-Dev3} \\
& \Lnot{(A \Lolly B)} \Lolly \bot				& \using{From (\ref{thm-lolly-not-not-Dev3}) and Corollary \ref{thm139}} \label{thm-lolly-not-not-Dev4} \\
& \Lnot{\Lnot{(A \Lolly B)}} \hspace{35mm}		& \using{By (\ref{thm-lolly-not-not-Dev4})} \tag*{\Done}
\end{align}

%%%%%%%%%%%%%%%%%%%%%%%%%%%%%%%%%%%%%%%%%%%%%%%%%%
\subsection{Double negation homomorphism: Conjunction}
%%%%%%%%%%%%%%%%%%%%%%%%%%%%%%%%%%%%%%%%%%%%%%%%%%
\label{homo-dn-tensor}

As done in Section \ref{homo-dn-lolly} for implication, we now show that (in $\LLi$) the double negation mapping $A \mapsto \Lnot{\Lnot{A}}$ is also a homomorphism for conjunction, i.e.
\[ \Lnot{\Lnot{(A \iAnd B)}} \;\Equiv{}\; \Lnot{\Lnot{A}} \iAnd \Lnot{\Lnot{B}}. \]
This result will follow immediately from a duality between implication ($\Lolly$) and conjunction ($\iAnd$) -- Theorem \ref{lolly-plus} below.

\begin{Lemma}[$\LLi$] \label{guess-il} $\Lnot{A} \Equiv{} (A \Lolly B) \iAnd \Lnot{(\CCb{A}{B})}$
\end{Lemma}
\Proof Left-to-right follows directly from Lemma \ref{guess}, taking $C =~\F$. For the converse observe that $\Lnot{(\CCb{A}{B})} \Equiv{} A \Lolly \Lnot{(A \Lolly B)}$. \Done

\begin{Theorem}[$\LLi$] \label{lolly-plus} $\Lnot{(\Lnot{A} \Lolly B)} \Equiv{} \Lnot{A} \iAnd \Lnot{B}$
\end{Theorem}
\Proof The implication from right to left is easy. For the other direction:
\begin{align}
& \Lnot{(\Lnot{A} \Lolly B)}					& \using{Given} \label{lolly-plus-Given1} \\
& (\CMb{\Lnot{\Lnot{A}}}{\Lnot{(\Lnot{A} \Lolly B)}}) \Lolly \Lnot{A} & \using{Derivable} \label{lolly-plus-Dev0} \\
& (\Lnot{A} \Lolly B) \Lolly \Lnot{\Lnot{A}}			& \using{By (\ref{lolly-plus-Given1}) and Lemma (\ref{guess-il})} \label{lolly-plus-Dev1} \\
& \Lnot{(\CCb{(\Lnot{A} \Lolly B)}{\Lnot{\Lnot{A}}})}	&  \label{lolly-plus-Dev5} \\
& \Lnot{(\CCb{\Lnot{A}}{B})}					& \using{By (\ref{lolly-plus-Dev1}), easy} \label{lolly-plus-Dev2} \\
%& \CMb{B}{\Lnot{\Lnot{A}}}					& \using{By (\ref{lolly-plus-Dev2}) and Theorem \ref{c-demorgan}} \label{lolly-plus-Dev3} \\
& \CMb{\Lnot{A}}{\Lnot{B}}					& \using{By (\ref{lolly-plus-Dev2}) and Theorem \ref{c-demorgan}} \label{lolly-plus-Dev4} \\
& \CMb{\Lnot{\Lnot{A}}}{\Lnot{(\Lnot{A} \Lolly B)}}	& \using{By (\ref{lolly-plus-Dev5}) and Theorem \ref{c-demorgan}} \label{lolly-plus-Dev6} \\
& \Lnot{A}								& \using{By (\ref{lolly-plus-Dev0}) and (\ref{lolly-plus-Dev6})} \label{lolly-plus-Dev8} \\
& \Lnot{A} \iAnd \Lnot{B} \hspace{45mm}		& \using{By (\ref{lolly-plus-Dev4}) and (\ref{lolly-plus-Dev8})} \tag*{\Done}
\end{align}

\begin{Theorem}[$\LLi$] \label{main-thm-n-trans} $\Lnot{\Lnot{(A \iAnd B)}} \Equiv{} \Lnot{\Lnot{A}} \iAnd \Lnot{\Lnot{B}}$
\end{Theorem}
\Proof By Theorem \ref{lolly-plus}, $\Lnot{(\Lnot{\Lnot{A}} \Lolly \Lnot{B})} \Equiv{} \Lnot{\Lnot{A}} \iAnd \Lnot{\Lnot{B}}$. But it is easy to show that $\Lnot{(\Lnot{\Lnot{A}} \Lolly \Lnot{B})} \Equiv{} \Lnot{\Lnot{(A \iAnd B)}}$, even in $\ALi$ (cf. Lemma \ref{ali-basic-lemma}). \Done

%%%%%%%%%%%%%%%%%%%%%%%%%%%%%%%%%%%%%%%%%%%%%%%%%%
\Subsection{Some De Morgan Dualities for $\LLi$}
%%%%%%%%%%%%%%%%%%%%%%%%%%%%%%%%%%%%%%%%%%%%%%%%%%
\label{sec-de-morgan}

% The proof of the homomorphism property for $A \iAnd B$ (Theorem \ref{main-thm-n-trans}) we made essential use of 

Theorem \ref{c-demorgan} proves an interesting De Morgan duality between $\CC$ and $\CM$. For completeness, we now give analogous dualities for all of our connectives (primitive and derived). 
% Let us conclude the paper with a list of De Morgan laws (derivable in $\LLi$) for formulas of the form $\Lnot{(A \circ B)}$ with $\circ$ ranging over each of our connectives (primitive and derived). 
% For conjunction ($\iAnd$) this is a trivial consequence of (un)currying, whereas for the weak conjunction ($\CC$) this is shown in Theorem \ref{c-demorgan}. We prove here similar results for the other connectives.

\begin{Theorem} \label{thm-main-demorgan} The following De Morgan dualities hold in $\LLi$
\[
\begin{array}{rcl}
\Lnot{(A \iAnd B)} & \Equiv{} & A \Lolly \Lnot{B} \\[1mm]
\Lnot{(A \Lolly B)} & \Equiv{} & \Lnot{\Lnot{A}} \iAnd \Lnot{B} \\[1mm]
\Lnot{(\CCb{A}{B})} & \Equiv{} & \CMb{A}{\Lnot{B}} \\[1mm]
\Lnot{(\CMb{A}{B})} & \Equiv{} & \CCb{\Lnot{\Lnot{A}}}{\Lnot{B}} \\[1mm]
\Lnot{(\CCb{A}{B})} & \Equiv{} & \CKb{\Lnot{A}}{\Lnot{B}} \\[1mm]
\Lnot{(\CKb{A}{B})} & \Equiv{} & \CCb{\Lnot{A}}{\Lnot{B}} \\[1mm]
\Lnot{(\oCKb{A}{B})} & \Equiv{} & \CMb{\Lnot{A}}{\Lnot{\Lnot{B}}}
\end{array}
\]
\end{Theorem}
\Proof The first equation $\Lnot{(A \iAnd B)} \Equiv{} A \Lolly \Lnot{B}$ follows directly from currying and uncurrying. For the second equation we calculate as follows
\begin{align*}
\Lnot{(A \Lolly B)} & \Equiv{} \Lnot{\Lnot{\Lnot{(A \Lolly B)}}} \by{easy} \\[1mm]
	& \Equiv{} (\Lnot{\Lnot{A}} \Lolly \Lnot{\Lnot{\Lnot{B}})} \by{Theorem \ref{thm-lolly-not-not}} \\[1mm]
	& \Equiv{} \Lnot{(\Lnot{B} \Lolly \Lnot{A})} \by{easy} \\[1mm]
	& \Equiv{} \Lnot{\Lnot{A}} \iAnd \Lnot{B}. \by{Theorem \ref{lolly-plus}}
\end{align*}
The third equation follows from Theorem \ref{c-demorgan} and $\CWC$. The fourth equation can be derived as:
\begin{align*}
\Lnot{(\CMb{A}{B})}
	& \Equiv{} \Lnot{(A \Lolly A \iAnd B)} \by{def $\CM$} \\[1mm]
	& \Equiv{} \Lnot{\Lnot{A}} \iAnd \Lnot{(A \iAnd B)} \by{duality of $\Lolly$} \displaybreak[1] \\[1mm]
	& \Equiv{} \Lnot{\Lnot{A}} \iAnd (B \Lolly \Lnot{A}) \by{duality of $\iAnd$} \displaybreak[1] \\[1mm]
	& \Equiv{} \Lnot{\Lnot{A}} \iAnd (B \Lolly \Lnot{\Lnot{\Lnot{A}}}) \by{$\Lnot{A} \Equiv{} \Lnot{\Lnot{\Lnot{A}}}$} \\[1mm]
	& \Equiv{} \Lnot{\Lnot{A}} \iAnd (\Lnot{\Lnot{A}} \Lolly \Lnot{B}) \by{easy} \\[1mm]
	& \Equiv{} \CCb{\Lnot{\Lnot{A}}}{\Lnot{B}}. \by{def $\CC$}
\end{align*}
The fifth equation follows by:
\begin{align*}
\Lnot{(\CCb{A}{B})}
	& \Equiv{} \Lnot{(A \iAnd (A \Lolly B))} \by{easy} \\[1mm]
	& \Equiv{} \Lnot{\Lnot{(A \Lolly B)}} \Lolly \Lnot{A} \by{easy} \\[1mm]
	& \Equiv{} (\Lnot{\Lnot{A}} \Lolly \Lnot{\Lnot{B}}) \Lolly \Lnot{A} \by{Theorems \ref{thm-lolly-not-not}} \\[1mm]
	& \Equiv{} (\Lnot{B} \Lolly \Lnot{A}) \Lolly \Lnot{A} \by{easy} \\[1mm]
	& \Equiv{} \CKb{\Lnot{A}}{\Lnot{B}}.
\end{align*}
For the sixth equation we proceed as follows:
\begin{align*}
\Lnot{(\CKb{B}{A})}
	& \Equiv{} \Lnot{((A \Lolly B) \Lolly B)} \by{def $\CK$} \\[1mm]
	& \Equiv{} \Lnot{\Lnot{(A \Lolly B)}} \iAnd \Lnot{B} \by{duality of $\Lolly$} \\[1mm]
	& \Equiv{} \Lnot{(\Lnot{\Lnot{A}} \iAnd \Lnot{B})} \iAnd \Lnot{B} \by{duality of $\Lolly$} \\[1mm]
	& \Equiv{} (\Lnot{B} \Lolly \Lnot{\Lnot{\Lnot{A}}}) \iAnd \Lnot{B} \by{duality of $\iAnd$} \\[1mm]
	& \Equiv{} (\Lnot{B} \Lolly \Lnot{A}) \iAnd \Lnot{B} \by{$\Lnot{A} \Equiv{} \Lnot{\Lnot{\Lnot{A}}}$} \\[1mm]
	& \Equiv{} \CCb{\Lnot{B}}{\Lnot{A}}. \by{def $\CC$}
\end{align*}
The last equation follows from Theorem \ref{k-negated} and the laws for $\CC$ and $\CK$. \Done

%%%%%%%%%%%%%%%%%%%%%%%%%%%%%%%%%%%%%%%%%%%%%%%%%%
%%%%%%%%%%%%%%%%%%%%%%%%%%%%%%%%%%%%%%%%%%%%%%%%%%
\Section{Negative Translations of {\L}ukasiewicz Logic}
\label{sec:embedding}
%%%%%%%%%%%%%%%%%%%%%%%%%%%%%%%%%%%%%%%%%%%%%%%%%%
%%%%%%%%%%%%%%%%%%%%%%%%%%%%%%%%%%%%%%%%%%%%%%%%%%

In this section we show that all four translations considered (Kolmogorov, G\"odel, Gentzen and Glivenko) are negative translations for $\LLi$. In fact, as it is the case in $\IL$, it turns out that any two negative translations for $\LLi$ are equivalent.
%
% in the setting of {\L}ukasiewicz logic. Our original expectation was that the use of contraction in the proof of correctness for these translations was essential and that they would fail to meet the requirements on a translation of $\LLc$ into $\LLi$. It was a pleasant and interesting surprise to discover that all of the well-known negative translations can be proved correct using only the weak form of contraction available in {\L}ukasiewicz Logic. 
%
This is a non-trivial result, since, as we have shown, the Gentzen and Glivenko translations fail for $\ALi$. The crucial property we will need here is that the double negation mapping $A \mapsto \Lnot{\Lnot{A}}$ is a homomorphism in $\LLi$, as proven in Sections \ref{homo-dn-lolly} and \ref{homo-dn-tensor}.

% In Theorem \ref{thm-kolm} we have seen that Kolmogorov and G\"odel translations are negative translations for $\ALi$. Since in both cases the translation of $\CWC$ follows from $\CWC$ itself, we can conclude that Kolmogorov and G\"odel are also negative translations for $\LLi$. We now prove that the same holds for Glivenko and Gentzen translations.

\begin{Theorem} \label{theorem-glivenko} The Glivenko translation $\glTrans{(\cdot)}$ is a negative translation for $\LLi$.
\end{Theorem}
\Proof We show by induction on the structure of $A$ that $\kTrans{A} \Equiv{} \Lnot{\Lnot{A}}$ in $\LLi$. This is similar to the proof of Theorem \ref{thm-kolm}, where we showed that G\"odel's translation is equivalent to Kolmogorov's in $\ALi$. In here we need a slightly stronger version of Lemma \ref{ali-basic-lemma} ($vi$) which in fact follows from Theorem \ref{thm-lolly-not-not}. Again, the base case is trivial. For implication we have
\begin{align*}
\kTrans{(A \Lolly B)} & \Equiv{} \Lnot{\Lnot{(\kTrans{A} \Lolly \kTrans{B})}} \by{def $\kTrans{(\cdot)}$} \\[1mm]
	& \Equiv{} \Lnot{\Lnot{(\Lnot{\Lnot{A}} \Lolly \Lnot{\Lnot{B}})}} \by{IH} \\[1mm]
	& \Equiv{} \Lnot{\Lnot{\Lnot{\Lnot{(A \Lolly B)}}}} \by{Theorem \ref{thm-lolly-not-not}} \\[1mm]
	& \Equiv{} \Lnot{\Lnot{(A \Lolly B)}}. \by{easy}
\end{align*}
Similarly for conjunction
\begin{align*}
\kTrans{(A \iAnd B)} & \Equiv{} \Lnot{\Lnot{(\kTrans{A} \iAnd \kTrans{B})}} \by{def $\kTrans{(\cdot)}$} \\[1mm]
	& \Equiv{} \Lnot{\Lnot{(\Lnot{\Lnot{A}} \iAnd \Lnot{\Lnot{B}})}} \by{IH} \\[1mm]
	& \Equiv{} \Lnot{\Lnot{\Lnot{\Lnot{(A \iAnd B)}}}} \by{Lemma \ref{ali-basic-lemma} ($vi$)} \\[1mm]
	& \Equiv{} \Lnot{\Lnot{(A \iAnd B)}}. \by{easy}
\end{align*}
\Done 

But note that we have not yet used the full strength of our homomorphism properties for double negation, as we only used it in a ``negated context". We will make use of them now to show that any translation for $\LLi$ which satisfies {\bf (NT1)} and {\bf (NT2)} will in fact also satisfy {\bf (NT3)}. 

\begin{Lemma} \label{lemma-negative} For any formula $A \in \Negative$ we have that $A \Equiv{\LLi} \Lnot{\Lnot{A}}$.
\end{Lemma}
\Proof By induction on $A \in \Negative$. We need to consider three cases: \\[1mm]
If $A =~\F$ the result is trivial. \\[1mm]
If $A = B \Lolly C$ with $C \in \Negative$ then, by the inductive hypothesis, $C \Equiv{} \Lnot{\Lnot{C}}$. Hence $B \Lolly C \Equiv{} B \Lolly \Lnot{\Lnot{C}}$. Since $B \Lolly \Lnot{\Lnot{C}} \Equiv{} \Lnot{\Lnot{B}} \Lolly \Lnot{\Lnot{C}}$, even in $\ALi$, by Theorem \ref{thm-lolly-not-not} we have that $B \Lolly C \Equiv{} \Lnot{\Lnot{(B \Lolly C)}}$. \\[1mm]
If $A = B \iAnd C$ with $B, C \in \Negative$ then, by the inductive hypothesis, we have that $B \Equiv{} \Lnot{\Lnot{B}}$ and $B \Equiv{} \Lnot{\Lnot{B}}$, hence $B \iAnd C \Equiv{} \Lnot{\Lnot{(B \iAnd C)}}$ by Theorem \ref{main-thm-n-trans}. %\\[1mm]
\Done

\begin{Theorem} \label{theorem-main} Any translation $(\cdot)^\dagger$ for $\LLi$ which satisfies {\bf (NT1)} and {\bf (NT2)} is equivalent to $\glTrans{(\cdot)}$ and hence is a negative translation, i.e., $(\cdot)^{\dagger}$ also satisfies {\bf (NT3)}.
\end{Theorem}
\Proof Fix a formula $A$. By {\bf (NT1)}, $A^\dagger \Equiv{} B$ with $B \in \Negative$. That $A^\dagger \Equiv{\LLi} \glTrans{A}$ can be shown as
\begin{align*}
A^\dagger & \Equiv{} B \\[1mm]
	& \Equiv{} \Lnot{\Lnot{B}} \tag{Lemma \ref{lemma-negative}} \\[1mm]
	& \Equiv{} \Lnot{\Lnot{A^\dagger}} \tag{since $A^\dagger \Equiv{} B$} \\[1mm]
	& \Equiv{} \Lnot{\Lnot{A}} \tag{by {\bf (NT2)} and Theorem \ref{theorem-glivenko}}.
\end{align*}
By Theorem \ref{theorem-glivenko}, $\glTrans{(\cdot)}$ satisfies {\bf (NT3)}, hence so does $(\cdot)^{\dagger}$.
\Done

\begin{Corollary} \label{gentzen-soundness} The Gentzen translation $\ggTrans{(\cdot)}$ is a negative translation for $\LLi$.
\end{Corollary}
\Proof Since $\ggTrans{(\cdot)}$ satisfies {\bf (NT1)} and {\bf (NT2)} in $\LLi$.
\Done \\

Theorem \ref{theorem-main} can be used to conclude that several other formula translations are also negative translations for $\LLi$. 

\begin{Example} Define a variant of the G\"odel translation whereby the definition of $\godTransPre{(A \Lolly B)}$ is modified as
\[ \godTransPre{(A \Lolly B)} \equiv A \Lolly \Lnot{\Lnot{\godTransPre{B}}}, \]
i.e. the premise of the implication is not inductively translated. It is easy to see that this ``simplification" still satisfies {\bf (NT1)} and {\bf (NT2)} and hence, by Theorem \ref{theorem-main}, is a negative translation for $\LLi$. A similar simplification can be considered for the Gentzen translation, leading, again, to a negative translation for $\LLi$.
\end{Example}

\begin{Example} Define $\krTransPre{A} \in {\cal L}$ inductively as follows:
\[
\begin{array}{rcl}
	\krTransPre{P} & \equiv & \Lnot{P} \quad\quad (\mbox{$P$ atomic}) \\[2mm]
	\krTransPre{\F} & \equiv & \T \\[2mm]
	\krTransPre{(A \iAnd B)} & \equiv & \krTransPre{A} \Lolly \Lnot{\krTransPre{B}} \\[2mm]
	\krTransPre{(A \Lolly B)} & \equiv & \krTransPre{A} \iAnd \Lnot{\krTransPre{B}}.
\end{array}
\]
Then we define the \emph{Krivine translation} of $A$ as $\krTrans{A} = \Lnot{\krTransPre{A}}$. The formula $\krTrans{A}$ is clearly a negative formula. It is also easy to check that $\krTrans{A} \Equiv{\LLc} A$. Therefore, by Theorem \ref{theorem-main}, it is a negative translation for $\LLi$. This translation is inspired by the negative translation behind Krivine's classical realizability interpretation \cite{Krivine(2003),Oliva(2008A)}.
\end{Example}

%Immediately from Theorem \ref{k-negated} and the last identity of Theorem \ref{thm-main-demorgan} it follows that $\oCKb{A}{B} \Equiv{\LLi} \CCb{\Lnot{A}}{\Lnot{B}}$. Unfolding the definitions of $\oCK$ and $\CC$ the corollary says 
%%
%\[ \Lnot{A} \iAnd (B \Lolly A) \Equiv{} \Lnot{A} \iAnd (\Lnot{A} \Lolly \Lnot{B}) \]
%%
%which means that \emph{in the context} $\Lnot{A} \iAnd (\cdot)$ the two implications $B \Lolly A$ and $\Lnot{A} \Lolly \Lnot{B}$ are intuitionistically equivalent. \\

\Section{Concluding Remarks}
\label{final-section}

Let us conclude with an argument that supports our choice of the name ``intuitionistic {\L}ukasiewicz logic" for the logic $\LLi$. First, an ``intuitionistic {\L}ukasiewicz logic" should be both a fragment of {\L}ukasiewicz logic $\LLc$ and intuitionistic logic $\IL$; and $\LLi$ satisfies this criteria. But there are indeed other "logics" which also satisfy this criteria, so why to single out $\LLi$? 

First, one might try to simply take the intersection of $\IL$ and $\LLc$. This is indeed the maximal set of logical theorems which are both valid in $\LLc$ and $\IL$. But it is not clear to us how one could give a simple sequent calculus for such logic. Moreover, this logic would not have the disjunction property, since $\DNE \vee \CON$ belongs to the intersection, but $\DNE$ is not provable in $\IL$, and $\CON$ is not provable in $\LLc$. Indeed, this has gone beyond what a constructivist would accept as an ``intuitionistic" fragment of $\LLc$. 

Given our results above about the soundness of the negative translations for $\LLc$ and $\LLi$, we want to argue that $\LLi$ is the only logic to be an extension of $\ALi$ having this soundness property. More precisely, let $\Delta$ be some axiom schema such that
\begin{itemize}
	\item[(1)] $\ALi + \Delta \subset \IL$, i.e. $\Delta$ is intuitionistically valid,
	\item[(2)] $\ALi + \Delta + \DNE = \LLc$, i.e. $\Delta$ and $\DNE$ together takes $\ALi$ to $\LLc$
\end{itemize}
and suppose the Glivenko translation is a negative translation for $\ALi + \Delta$, i.e.
\begin{description}
	\item[(3)] if $\LLc$ proves $\vdash A$ then $\ALi + \Delta$ proves $\vdash \Lnot{\Lnot{A}}$.
\end{description}
We argue that over $\ALi$ we have $\Lnot{\Lnot{\Delta}} \Equiv{} \Lnot{\Lnot{\CWC}}$, in the sense that for any instance $B$ of $\CWC$ there exists $A_1, \ldots, A_n \in \Delta$ such that $\ALi$ proves $\Lnot{\Lnot{A_1}}, \ldots, \Lnot{\Lnot{A_n}} \vdash \Lnot{\Lnot{B}}$; and similarly with $\CWC$ and $\Delta$ interchanged. 

First, assume $B \in \CWC$. Since $\LLc$ proves $\vdash B$, by (3) we have that $\ALi + \Delta$ proves $\vdash \Lnot{\Lnot{B}}$. Which means that for some $A_1, \ldots, A_n \in \Delta$ we have that $\ALi$ proves $\Lnot{\Lnot{A_1}}, \ldots, \Lnot{\Lnot{A_n}} \vdash \Lnot{\Lnot{B}}$, using Lemma \ref{ali-basic-lemma} ($vi$). 

Conversely, assume $B \in \Delta$. Again, since $\LLc$ proves $\vdash B$, by our soundness result for the Glivenko translation (Theorem \ref{theorem-glivenko}), we have that $\ALi + \CWC$ proves $\vdash \Lnot{\Lnot{B}}$. Which means that for some $A_1, \ldots, A_n \in \CWC$ we have that $\ALi$ proves $\Lnot{\Lnot{A_1}}, \ldots, \Lnot{\Lnot{A_n}} \vdash \Lnot{\Lnot{B}}$, again using Lemma \ref{ali-basic-lemma} ($vi$). \\

\noindent {\bf Acknowledgements}. We are indebted to the late Franco Montagna for bringing \cite{Bova:2009} to our attention, and for a manuscript detailing how their decision procedure for commutative GBL-algebras leads to a decision procedure for $\LLi$. We would also like to thank George Metcalfe and Isabel Ferreirim for helpful correspondence.

%\begin{acknowledgements}
%\end{acknowledgements}

\bibliographystyle{spmpsci}      % mathematics and physical sciences

\bibliography{references}

\end{document}